\documentclass[preprint,12pt]{elsarticle}

\usepackage{amssymb}
\usepackage{amsmath}

\usepackage{cite}
\usepackage{amsmath,amssymb,amsfonts}
\usepackage{algorithmic}
\usepackage{graphicx}
\usepackage{textcomp}
\usepackage{xcolor}
\usepackage{algorithm2e}
\newtheorem{definition}{Definition}

\begin{document}

\begin{frontmatter}



\title{Improved Convolution-Based Analysis for Worst-Case Probability Response Time of CAN} 

\author[label1]{Haozhe Yi}
\affiliation[label1]{organization={University of Electronic Science and Technology of China},
             city={Chengdu},
             country={China}}

\author[label2]{Junyi Liu} 
\affiliation[label2]{organization={Northeastern University},
            city={Shenyang},
            country={China}}

\author[label1,label3]{Maolin Yang}
\fntext[label3]{Corresponding author.}

\author[label1]{Zewei Chen}

\author[label1]{Xu Jiang}

\begin{abstract}
Controller Area Networks (CANs) are widely adopted in real-time automotive control and are increasingly standard in factory automation. Considering their critical application in safety-critical systems, The error rate of the system must be accurately predicted and guaranteed.  Through simulation, it is possible to obtain a low-precision overview of the system's behavior.  However, for low-probability events, the required number of samples in simulation increases rapidly, making it difficult to conduct a sufficient number of simulations in practical applications, and the statistical results may deviate from the actual outcomes.  Therefore, a formal analysis is needed to evaluate the error rate of the system. This paper improves the worst-case probability response time analysis by using convolution-based busy-window and backlog techniques under the error retransmission protocol of CANs. Empirical analysis shows that the proposed method improves upon existing methods in terms of accuracy and efficiency.

\end{abstract}



\end{frontmatter}



\section{Introduction}
\label{sec1}

Probabilistic real-time analysis has become an important method for analyzing variable task execution times in real-time systems.  
Unlike traditional deterministic analysis, which assumes a single worst-case execution time for each task, probabilistic analysis leverages probability distributions to model the execution time of tasks.
Typically, the worst-case execution time of tasks has a very low probability of occurring, making it difficult to accurately describe the actual system when analyzing based on worst-case execution time. 
Probabilistic real-time analysis provides a more accurate representation of task behavior by considering the complex software and hardware interactions prevalent in modern real-time systems. 
A significant advantage of probabilistic analysis is its shift from guaranteeing absolute certainty in meeting timing requirements to evaluating the likelihood of meeting those requirements while allowing for the possibility of not meeting the requirements to some extent. By quantifying the probabilities of meeting deadlines, system designers can make informed decisions based on predefined limits \citep{bozhko2023}.

Many real-time applications, including critical systems like active steering in automotive contexts, operate within safety-critical environments where adherence to stringent safety standards is paramount. For example, the automotive standard ISO-26262  \citep{iso2000} specifies specific failure rates for each Automotive Safety Integrity Level (ASIL), providing a framework for decision-making in the automotive industry. 
In the context of Controller Area Networks (CAN), which are extensively used in automotive and industrial applications, the consideration of errors and their implications is particularly crucial. Systems utilizing CAN protocols often incorporate fault-tolerance mechanisms to ensure reliability. However, these mechanisms introduce additional overhead in response to errors, such as increased signaling and recovery processes like Automatic Repeat Requests (ARQ), which initiate retransmissions. This overhead is essential for maintaining system integrity and meeting real-time constraints in the presence of unpredictable error events typical in CAN-based environments.

In this paper, we analyze the probabilistic guarantees on worst-case response times for the CAN protocol with error retransmission mechanisms and propose an improved reliable convolution-based algorithm to calculate the upper bound of the probability that response times exceed a limit. 

The contributions of this paper are described as follows:
\begin{itemize}
    \item We construct a probability CAN model with a probabilistic error retransmission mechanism.
    \item We propose a novel  convolution algorithm based on busy-window and backlog to analyze the probabilistic worst-case response time for frames in CAN. Our algorithm provides an efficient convolution process and explicitly specifies the stopping conditions.
    \item Through extensive empirical evaluation encompassing a wide range of parameter settings, we demonstrate
that our new technique significantly outperforms state-of-theart methods in terms of analysis precision and speed.
\end{itemize}

The remainder of this paper is structured as follows: After summarizing related work in Section \ref{sec2}, Section \ref{sec3} presents the system model, including the CAN model and the error retransmission model, and introduces relevant probabilistic operation. Section 4 \ref{sec4}introduces our response time analysis method and convolution algorithm. Section 5 \ref{sec5} conducts comparative experiments. Finally, we conclude the paper and propose future work in Section \ref{sec6}.

\section{Related Works}
\label{sec2}

Recent research on probabilistic real-time systems has gained significant attention. Probabilistic scheduling in CAN bus systems involves analyzing the likelihood that tasks meet their deadlines while considering uncertainties in task execution times.

\subsection{Probabilistic Schedulability}

Systems are classified as real-time if they must satisfy both functional and timing requirements. Probabilistic real-time systems differ from the classical model in two key ways: firstly, they acknowledge that task execution times can vary significantly, and secondly, they do not demand absolute guarantees for meeting deadlines. Instead, they require that the probability of exceeding a deadline remains below a specified threshold.

In 1988, Woodbury and Shin proposed a probabilistic analysis method for periodic tasks \citep{Woodbury1988}, transforming tasks into paths with execution times and occurrence probabilities. They iteratively calculated the remaining execution time of higher-priority jobs and the probability distribution of release job execution times, thereby determining the probability of all jobs missing their deadlines within the hyperperiod.

In 1995, Tia et al. proposed the PTDA algorithm, which uses convolution to calculate the probability distribution of the cumulative execution time of released tasks at each scheduling point under fixed-priority preemptive scheduling \citep{Tia1995}.

Subsequent research has explored issues related to identifying worst-case instance \citep{Gardner1999}, the connection between reliable results and the concept of pessimism \citep{Diaz2004}, and the impact of execution time dependencies among tasks on analysis \citep{Ivers2009}. Analyses have been extended to continuous variable distributions \citep{Tanasa2015}, cases involving backlog \citep{diaz2002}, and queue jitter \citep{Tanasa2013}. Additionally, various algorithms for probabilistic scheduling have been proposed.

\subsection{Analysis of CAN system}

CAN’s successful history in the automotive industry, as a versatile embedded bus, has led to its adoption in various applications including factory-control networks, elevator systems, machine tools, and photocopiers. However, for safety-critical avionics applications, CAN is still regarded as insufficiently reliable and predictable. This has led to increased interest in analyzing CAN’s real-time performance.

Tindell and Burns (1994) \citep{tindell1994} developed a schedulability analysis for CAN under deterministic conditions. They assessed the worst-case response time of a given hard real-time message by considering queuing jitter, worst-case queuing delay, and the longest time taken to send a message.

To address error retransmissions, a 2002 study proposed a probabilistic tree-based analysis that iteratively calculates error probabilities over non-overlapping intervals. However, this approach assumes each error results in the transmission of the longest erroneous frame and the retransmission of the longest high-priority frame, introducing additional pessimism. It also fails to account for response times exceeding the message period, making it unable to guarantee the worst-case response time \citep{broster2002}.

A 2013 study introduced a convolution-based analysis model using the Stochastic Busy Window. This model utilizes execution time probability mass functions (pmfs) to account for variable execution time overhead caused by error events. Despite its innovative approach, the method does not explicitly specify the stopping condition for the queuing delay iteration. As a result, it is impossible to determine the queuing delay corresponding to the response time of the \(q\)-th event. Additionally, the queuing delay pmf for each instance needs to iterate from the critical instant, which is unnecessary \citep{axer2013}.

Moreover, in a CAN bus system, all nodes need to be strictly synchronized. Therefore, once the Start of Frame (SOF) bit begins to transmit, any node that has not yet sent its message cannot enter the current arbitration round until the data transmission is completed and the bus returns to an idle state. Consequently, any frame that arrives at the end of the busy window will not be able to join the current arbitration round, thus not interfering with the busy window calculation. 

Another study from 2002 proposed an analysis method for the response time distribution of preemptive periodic real-time systems. This method calculates the probability distribution of the backlog at task release times, considering the remaining execution times that have not yet been completed. It iterates by convoluting the new tasks, ultimately computing the backlog probability distribution matrix at the start of hyperperiods and demonstrating that this process is a Markov process \citep{diaz2002}. 

To analyze the worst-case scenario, we do not need to analyze multiple hyperperiods but can iterate from the critical instant, and in systems with utilization less than 100\%, the iteration will necessarily terminate in a finite amount of time. Moreover, for the case of non-preemptive fixed-priority scheduling, we must also account for blocking caused by lower-priority tasks and delays caused by continuous preemption by higher-priority tasks.

\section{System Model and Notation} 
\label{sec3}
In this section, we describe the CAN model and define some concepts used in our analysis. We assume a discrete-time model, where the variable $\Delta t$ represents an indivisible unit of time, i.e., the time required to send a single bit.

\subsection{CAN Model}

Controller Area Network(CAN) CAN is a serial data bus that supports priority based message arbitration and non-preemptive message transmission. 
Each CAN data frame includes an identifier, and the CAN protocol mandates that nodes wait for a bus idle period before attempting to transmit. If two or more nodes begin transmitting simultaneously, each node monitors each bit on the bus to ascertain whether it is transmitting the highest priority message (indicated by a numerically lower identifier) and should continue, or if it should stop transmitting and wait for the next bus idle period to try again.

The system can be modeled as a message set consisting of $n$ independent sporadic frames $F = \{\tau_0, \tau_1, \ldots, \tau_n\}$, transmitted on the bus according to fixed priority non-preemptive scheduling. Each CAN frame is characterized by the tuple $(C_i, T_i, D_i, p_i)$:

\begin{itemize}
    \item \textbf{Worst-case transmission time ($C$)}: The maximum time it takes to transmit the entire frame's payload.
    \item \textbf{Minimum inter-arrival time ($T$)}: The minimum time interval required between successive arrivals of the same frame type.
    \item \textbf{Relative deadline ($D$)}: The time at which the frame must be successfully transmitted after it is released.
    \item \textbf{Fixed priority ($p$)}: A numerical value determining the frame's priority on the bus, with lower values indicating higher priority.
\end{itemize}

Next, we define $\tau_i$ has higher priority than frame \(\tau_j\) if \(i<j\). We denote $hp(\tau_i)$ as the set of frames with higher priority than $\tau_i$.

\subsection{Error Retransmission Model}

CAN was designed as a robust and reliable form of communication for short messages. Each frame has a 15-bit Cyclic Redundancy Check (CRC), which is used by receiving nodes to check for errors in the transmitted message. If an error is detected, the node will send an error frame as a signal and a new arbitration phase will begin after the frame. This means that in case of an error event, the error signaling causes additional time overhead, and previously released higher priority instances are admitted after this signaling. We denote the worst-case error-signaling overhead as \( E \).

In this paper, we assume the use of an error retransmission mechanism. When a frame error is detected and an error signal is sent, the node will attempt retransmission for up to \(C_i\) units of time. All transmissions, including initial transmission and retransmissions, are treated as part of the same instance. An error may occur at any time during the instance, but it will not interrupt the current transmission. This means that the frame must be transmitted in full before the error frame can be sent.

Next, we employ the single-bit-error model for the subsequent analysis. In this model, the probability that a bit is affected by an error remains constant and independent of previous errors and time. The occurrence of error events is modeled using a Poisson process with a specified error rate $\lambda$ \citep{broster2004}.

For the Poisson model, the following equations give the probability that exactly m error events occur (i.e. bit-flips), the probability that no error at all occurs and the converse probability that at least one error occurs during a given time interval $\delta$.

\[\mathbb{P}(X = m, \delta) = \frac{{(\lambda \delta)^m e^{-\lambda \delta}}}{{m!}}\]
\[\mathbb{P}^{ok}(\delta) = e^{-\lambda \delta}\]
\[\mathbb{P}^{errors}(\delta)) = 1 - e^{-\lambda \delta}\]

As shown in Fig.1, considering that the number of error events has a probability distribution, the total overhead of each instance needs to be considered using the probability mass function(pmf) rather than a fixed transmission time. This transformation converts the model into an equivalent variable transmission times problem.Considering that there is no upper limit on the number of retransmissions due to frame errors, for ease of analysis and practical considerations, we impose a maximum retry limit $k$. The value of $k$ is determined by the channel quality \( \lambda \) and the required transmission assurance.

\begin{definition}
   The variable transmission time $\mathcal{C}_i$ can be represented as the following pmf:
\[
\small
\begin{pmatrix}
C_i & 2C_i + E & \cdots & C_i + k(C_i + E) \\
\mathbb{P}(\mathcal{C}_i = C_i) & \mathbb{P}(\mathcal{C}_i = 2C_i + E) & \cdots & \mathbb{P}(\mathcal{C}_i = C_i + k(C_i + E))
\end{pmatrix}
\]
\normalsize 
\end{definition}

where $\mathbb{P}(\mathcal{C}_i=C_i+n(C_i+E))$ represents the probability of exactly $n$ error retries occurring, and it can be expressed as:

\[
\mathbb{P}_n = 
\begin{cases}
\mathbb{P}^{\text{ok}}(C) & \text{if } n = 0 \\
\mathbb{P}^{\text{errors}}(C) \cdot \left( \mathbb{P}^{\text{errors}}(C + E) \right)^{n-1} \cdot \mathbb{P}^{\text{ok}}(C + E) & \text{if } n > 0
\end{cases}
\]

Therefore, the upper bound on the cumulative probability of up to \( k \) retries can be expressed as \( \sum_{n=0}^{k} \mathbb{P}_n \), and the value of \( k \) can be determined by calculating the probability of exceeding \( k \) retries( \( 1 - \sum_{n=0}^{k} \mathbb{P}_n \)).

\begin{figure}[htbp]
    \centering
    \includegraphics[width=0.8\textwidth]{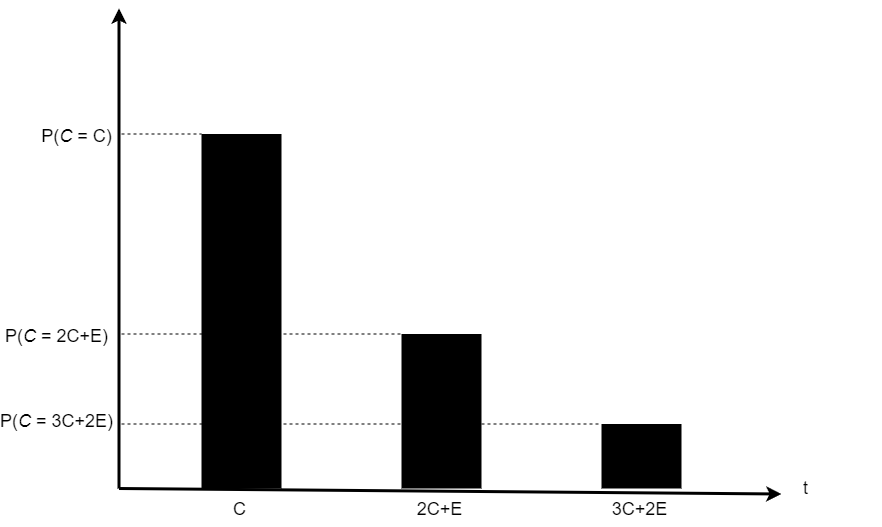}
    \caption{Worst-Case execution time pmf $\mathcal{C}$}
    \label{fig:your_label}
\end{figure}

We analyze the distribution of the probabilistic worst-case response time (pWCRT) and define the exceedance function, which represents the probability that the response time \( R_i \) exceeds a certain time \( t \). For all instances \(\tau_{i,j}\) of \(\tau_i\), we take the maximum value of probability where \( R > t \). The exceedance function is expressed as follows:

\[F_{R_i}(t) = \max_{j\in \mathbb{N}} \mathbb P(R_{i,j} > t)\]

The aim of this paper is to derive a safe (yet as precise as possible) upper bound of R$_i$.

In the CAN model, we describe each CAN frame using the tuple \((C_i, T_i, D_i, p_i)\). Subsequently, we need to provide some clarifications. According to error retransmission mechanisms, the worst-case transmission time of released instances of the frame should be represented by the random variable \(\mathcal{C}_i\). For sporadic frames, \(T_i\) denotes the minimum inter-arrival time, and we assume it satisfies the constrained deadline i.e. \(D_i\leq T_i\). 

\subsection{Definition of Probability Operations}

We begin by defining a random variable \( \mathcal{X} \) and its probability distribution \(\mathbb{P}( \mathcal{X}=x) \) \citep{chow2012}.

\begin{definition}
    A probability mass function (pmf) of a discrete random variable \( \mathcal{X}\) is a function \( f_X(x) \) that assigns probabilities to each possible value of \( \mathcal{X} \). The pmf \( f_\mathcal{X}(x) \) can be represented as:
\[f_\mathcal{X}(x)=
\begin{pmatrix}
x_1 & x_2 & \cdots & x_n \\
p_{x_1} & p_{x_2} & \cdots & p_{x_n} \\
\end{pmatrix}
\]
\end{definition}

Here, \( x_1, x_2, \ldots, x_n \) are the possible values that \( \mathcal{X} \) can take, and \( p_{x_i} = \mathbb{P}( \mathcal{X}=x_i) \) represents the probability associated with each value \( x_i \).

Next, we introduce the concept of coalescence.

\begin{definition}
   The coalescence \(  \mathcal{Z}=  \mathcal{X}\oplus \mathcal{Y} \) represents a combination of two partial random variables \(  \mathcal{X} \) and \( \mathcal{Y}\), defined as:
\[
\mathbb{P}( \mathcal{Z} = z) = \mathbb{P}( \mathcal{X} = z) + \mathbb{P}( \mathcal{Y} = z)
\]

\end{definition}

An example of coalescence is:

\[
\begin{pmatrix}
1 & 2 \\
0.2 & 0.4
\end{pmatrix}  \oplus
\begin{pmatrix}
2 & 4 \\
0.1 & 0.3
\end{pmatrix}=
\begin{pmatrix}
1 & 2 & 4 \\
0.2 & 0.5 & 0.3
\end{pmatrix}
\]

Next, we introduce the concept of convolution for independent discrete random variables.

\begin{definition}
  The \textbf{convolution} of two discrete random variables \( \mathcal{X} \) and \( \mathcal{Y} \) produces a new random variable \( \mathcal{Z} \), which represents the sum of \( \mathcal{X} \) and \( \mathcal{Y} \). The pmf of \( \mathcal{Z} \), denoted as \( f_Z(z) \), is given by:

\[f_Z(z) = \sum_{k} f_X(k) f_Y(z - k)\]   
\end{definition}

An example of convolution is:
\[
\begin{pmatrix}
3 & 4 \\
0.2 & 0.8
\end{pmatrix}  \otimes
\begin{pmatrix}
1 & 2 \\
0.3 & 0.7
\end{pmatrix}=
\begin{pmatrix}
4 & 5 & 6 \\
0.06 & 0.38 & 0.56
\end{pmatrix}
\]

\section{Analysis}
\label{sec4}
In this section, we conduct a deterministic analysis of CAN and review an algorithm that uses the busy-window technique to compute worst-case response times (WCRT). However, this algorithm is imperfect, and we improved it on the CAN model, proposing a new method for busy-window and backlog calculations.

\subsection{Deterministic Analysis}

The response time \(R_{i,j}\) is the time interval from the released of a instance \(\tau_{i,j}\) until the it has fully been transmitted. The worst-case response time for frame \(\tau_i\) occurs for some instances released within an \(i\)-level busy-window that starts immediately after the longest lower priority frame begins transmission in critical instant. An \(i\)-level busy-window is defined as:

\begin{itemize}
    \item It starts at some time \(t^s\) when instances of frames in \(\text{hp}(\tau_i) \cup \tau_i\) is released for transmission, and there are no instances of frames in \(\text{hp}(\tau_i) \cup \tau_i\) waiting to be transmitted that were released before the \(t^s\).
    \item It is a contiguous interval of time during which any instance of a frame with priority lower than \(i\) is unable to start transmission and win arbitration.
    \item It ends at the earliest time \(t^e\) when the bus becomes idle, ready for the next round of transmission and arbitration, yet there are no instances of frames in  \(\text{hp}(\tau_i) \cup \tau_i\)  waiting to be transmitted that were released before time \(t^e\).
\end{itemize}

A critical instant for a frame is defined to be an instant at which an instance for that frame will have the largest response time \citep{liu1973}. According to prior research, given a sporadic real-time frame system, the critical instant of a frame \(\tau_i\) under single-bus non-preemptive fixed-priority scheduling occurs at the release of a instance of \(\tau_i\) when:

\begin{itemize}
    \item Every higher-priority frame in \(\text{hp}(\tau_i)\) releases a job simultaneously with the instance of \(\tau_i\).
    \item All subsequent instances of the higher-priority tasks are released as early as possible by respecting their minimum inter-arrival times.
    \item Every instance is executed with its worst-case transmission time.
\end{itemize}

For non-preemptive scheduling, it is also necessary to consider the blocking caused by lower-priority frames in \(\text{lp}(\tau_i)\). The maximum blocking time \(B_i\) is given by the longest transmission time with error of the \(\text{lp}(\tau_i)\), represented as:

\[
B_i = \max_{\forall j \in \text{lp}(i)} (C_j+E),
\]

Therefore, by calculating the busy-window at the critical instant, including the blocking \(B_i\), we can determine the longest busy-window, which is then used to determine the worst-case response time \(R\) by \(C_i\). 

\subsection{Stochastic Analysis}

For the stochastic case, the Worst-Case Response Time (WCRT) can be calculated using the following formula:

\[
\mathcal R_{i,j} = \mathcal S_{i,j} + C_i
\]

The pmf of \( \mathcal S_{i,j} \) can be represented as:
\[
\begin{pmatrix}
s_{i,j,1} & s_{i,j,2} & \cdots & s_{i,j,n} \\
P(S_{i,j}=s_{i,j,1}) & P(S_{i,j}=s_{i,j,2}) & \cdots & P(S_{i,j}=s_{i,j,n})
\end{pmatrix}
\]

\(\mathcal S_{i,j}\) is the time from the release of \(\tau_{i,j}\) at \(r_{i,j}\) to the start of successful transmission without errors. \(r_{i,j}\) is the release time of the \(j\)-th instance of \(\tau_i\). \(S_{i,j}\) is computed as the convolution of the following components:

\begin{itemize}
  \item Backlog \( \mathcal B_{i,j} \) at Time \( r_{i,j} \)
  \item Error Retransmission Time \( \mathcal{C}_i^{\text{retrans}} \)
  \item Transmission Times of Higher-Priority Instances
  \item Transmission Time \( C_i \) of frame \( \tau_i \)
\end{itemize}

The backlog at time \( r_{i,j} \), denoted as \( \mathcal B_{i,j} \), is defined as the sum of the remaining transmission times of all instances released by \( \text{hp}(\tau_i) \cup \tau_i \) that have not completed transmission at time \( r_{i,j} \). This backlog \( \mathcal B_{i,j} \) can be derived using the busy-window approach. However, unlike the deterministic case, our busy-window is a probability distribution. As for the non-stochastic busy-window approach, the critical instant assumption remains valid also for the case where the transmission time is a random variable. We iteratively compute the busy-window according to the release order of instance from the critical instant.

The length of the busy-window is denoted as \(w_i^t\), representing the probability distribution of the duration of the \(i\)-level busy-window at the time t. The pmf of \(w_i^t\) can be represented as a matrix:

\[
\begin{pmatrix}
{W}_{i,t,1} & {W}_{i,t,2} & \cdots & {W}_{i,t,k} \\
\mathbb{P}(w_i^t = {W}_{i,t,1}) & \mathbb{P}(w_i^t = {W}_{i,t,2}) & \cdots & \mathbb{P}(w_i^t = {W}_{i,t,k})
\end{pmatrix}
\]

where \({W}_{i,t,k}\) represents the possible durations of the busy window, and \(\mathbb{P}(w_i^n = {W}_{i,t,k})\) represents their corresponding probabilities. Next, we will present the calculation method.

At the start of the analysis ($t=0$), the initial state of the $i$-level busy-window, denoted as \( w_{i}^0 \), is defined as:

\[ w_i^0 = \begin{pmatrix} B_i \\ 1 \end{pmatrix} \]

We assume \( w_i^{r_n} \) is known. Now, we need to compute \( w_i^{r_{n+1}} \) for \( r_{n+1} > r_n \), where \( r_n \) is the release time of the $n$-th instance of frame in \(\text{hp}(\tau_i) \cup \tau_i\). We sort the instances by their release times, and if the release times are the same, we sort by priority.

We decompose \( w_i^{r_n} \) into two parts: the stable part and the pending part. The stable part is defined as \( w_i^{r_n} \) in the interval \([0, t]\). The pending part is defined as \( w_i^{r_n} \) in the interval \((t, +\infty)\).

We define a splitting function \( f^{[a, b]}(x) \), which is defined as:

\[
f^{[a, b]}(x) = \begin{cases} 
x & \text{if } a \leq x \leq b, \\
0 & \text{otherwise}.
\end{cases}
\]

Using this function, we can decompose \( w_i^{r_n} \) as follows:

\[
w_i^{r_n} = f^{[0, t]}(w_i^{r_n}) \oplus f^{(t, +\infty)}(w_i^{r_n})
\]

When calculating \( w_i^{r_{n+1}} \), the new instance affects only the pending part. Therefore, \( w_i^{r_{n+1}} \) can be obtained by convolving the pending part with the instance released at \( r_n \), and then combining it with the stable part:

\[
w_i^{r_{n+1}} = f^{[0, t]}(w_i^{r_n}) \oplus \left( f^{(t, +\infty)}(w_i^{r_n}) \otimes \mathcal C_n \right)
\]
\[\forall t \in (r_n, r_{n+1}), \quad \text{no instances in \(\text{hp}(\tau_i) \cup \tau_i\) are released at } t.\]

where, \(\mathcal{C}_n\) denotes the probabilistic transmission time of instance \(n\).

The iteration of \( w_i^{r_n} \) stops when the probability that the next instance released by a frame in \( \text{hp}(\tau_i) \cup \tau_i \) falls within the busy-window is less than \( \epsilon \). At this point, it is determined that subsequent instances of \( \tau_{i,j} \) are outside the busy-window, thereby concluding the global computation since they will not contribute to the worst-case deadline miss probability. The threshold \( \epsilon \) is typically derived from a confidence interval.

This stopping condition can be expressed as:

\[
\sum_{W_{i,r_n,k} > r_n} \mathbb{P}(w_i^{r_n}=W_{i,r_n,k}) < \epsilon
\]

When calculating \( w_i^{r_{i,j}} \), if multiple \( \tau_n \) are released at \( r_{i,j} \), we choose \( w_i^{r_n} \) at the smallest \( n \), which is the busy window before the first instance is released at point \( r_{i,j} \).

\begin{figure*}[htbp]
    \centering
    \includegraphics[width=\textwidth]{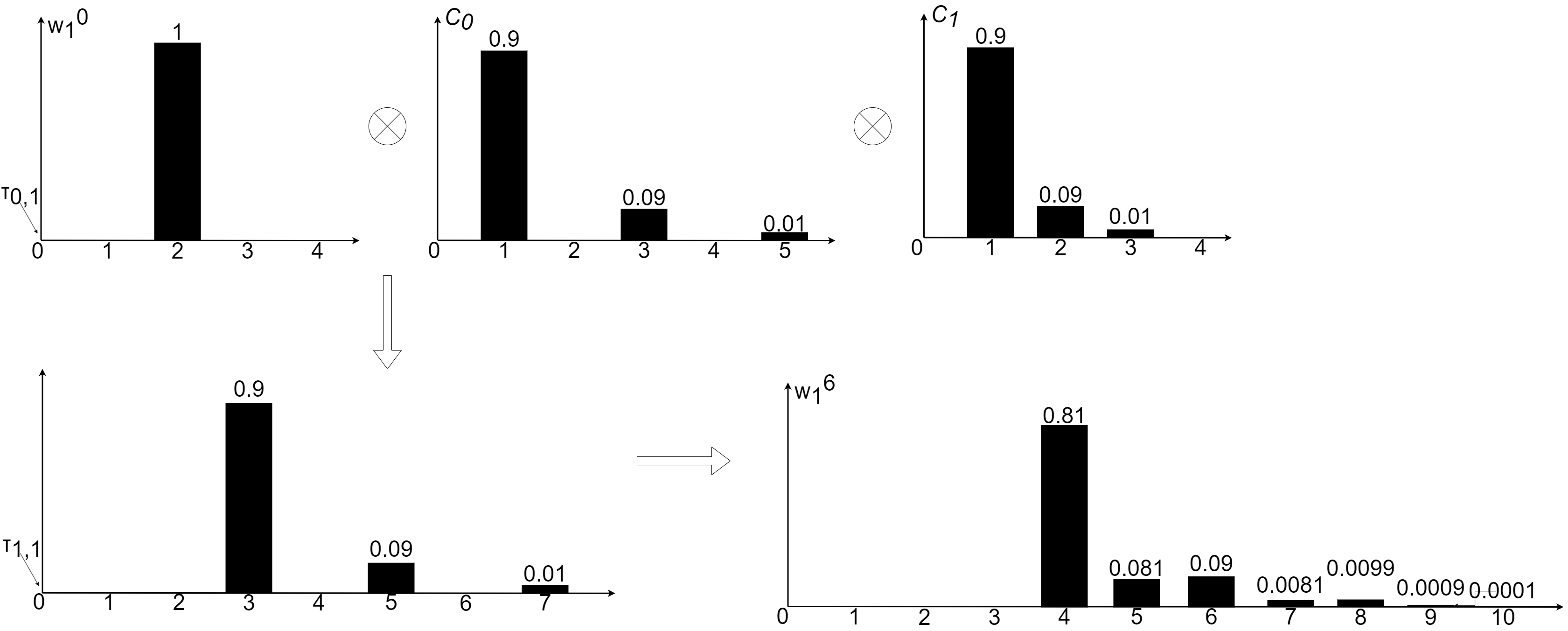}
    \caption{Example for constructing the busy-window(1)}
    \label{fig:your_label}
\end{figure*}

\begin{table}[h!]
    \centering
    \begin{minipage}[t]{0.45\textwidth}
        \centering
        \caption{Example frame's four-tuple}
        \begin{tabular}{|c|c|c|c|c|}
        \hline
        Frame & $C_i$ & $D_i$ & $T_i$ & Prio \\
        \hline
        $\tau_0$ & 1 & 6 & 6 & 0 \\
        $\tau_1$ & 1 & 12 & 12 & 1\\
        $\tau_2$ & 2 & 20 & 20 & 2\\
        \hline
        \end{tabular}
        \label{tab:four_tuple}
    \end{minipage}
    \hspace{0.05\textwidth} 
    \begin{minipage}[t]{0.45\textwidth}
        \centering
        \caption{Probability distribution of frame execution time}
        \begin{tabular}{|c|c|c|c|}
        \hline
        Frame & $C_i$ & $2C_i + E$ & $3C_i + 2E$ \\
        \hline
        $\tau_0$ ($\mathcal{C}$) & 1 & 3 & 5 \\
        $\tau_0$ (P) & 0.9 & 0.09 & 0.01 \\
        \hline
        $\tau_1$ ($\mathcal{C}$) & 1 & 2 & 3 \\
        $\tau_1$ (P) & 0.9 & 0.09 & 0.01 \\
        \hline
        \end{tabular}
        \label{tab:prob_dist}
    \end{minipage}
\end{table}

Fig.2 shows an example. In calculating the busy-window for $\tau_1$, it is blocked by $\tau_2$. Assuming $E$ for $\tau_2$ is 0, the maximum blocking duration $B_1$ is 2. Frames $\tau_0$ and $\tau_1$ belong to $\text{hp}(\tau_1) \cup \tau_1$. Instances of $\tau_0$ are released at times 0, 6, and 12. Instances of $\tau_1$ are released at times 0 and 12.Sorting the instances according to their release times and priorities, we get the instance queue:

\begin{table}[ht]
\centering
\caption{Instance queue sorted by release times and priorities.}
\begin{tabular}{|c|c|}
\hline
Time & Instances \\
\hline
0   & $\tau_{0,1}$, $\tau_{1,1}$ \\
6   & $\tau_{0,2}$ \\
12  & $\tau_{0,3}$, $\tau_{1,2}$ \\
\hline
\end{tabular}
\label{tab:instance_queue}
\end{table}

Assuming \(\epsilon = 0.00015\), the busy-window for \(w_{1}^{r_{0,1}}\) is initialized as \(B_i\). For \(w_{1}^{r_{0,2}}\), the instances \(\tau_{0,1}\) and \(\tau_{1,1}\) released before 6s are convolved sequentially to obtain \(w_{1}^{6}\). 

\begin{figure*}[htbp]
    \centering
    \includegraphics[width=\textwidth]{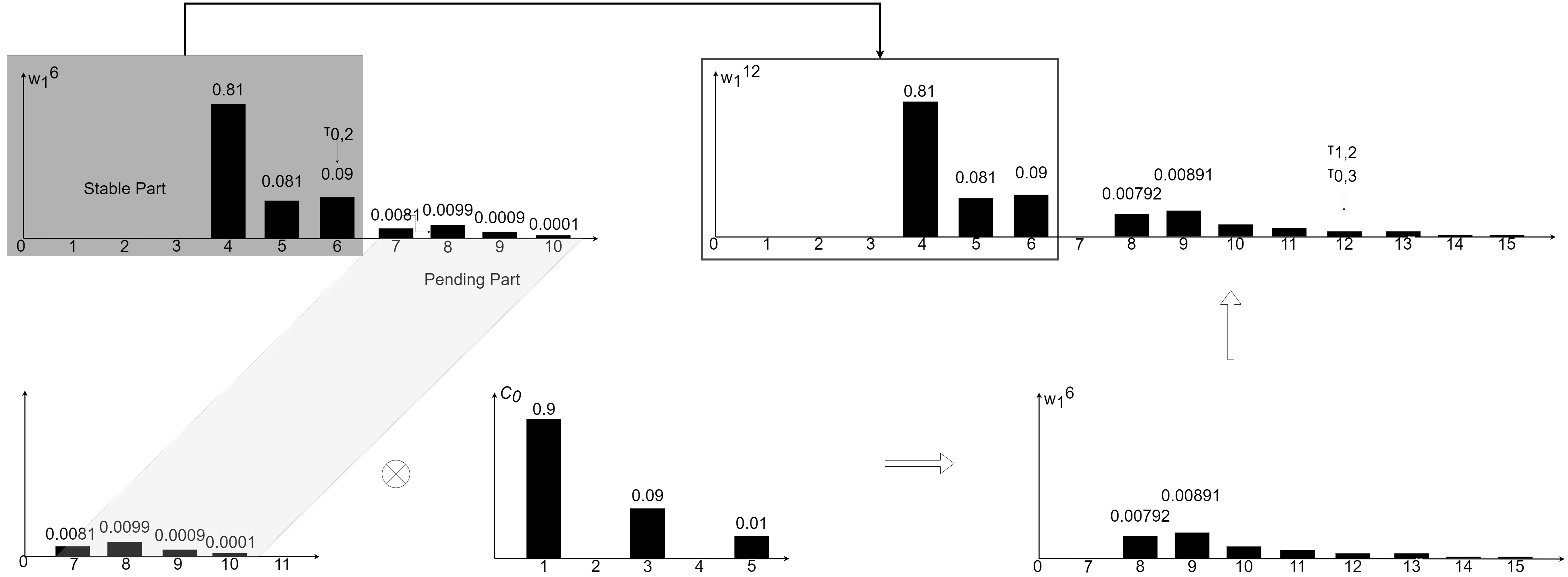}
    \caption{Example for constructing the busy-window(2)}
    \label{fig:your_label}
\end{figure*}

Fig.3 compute \(w_{1}^{r_{0,3}}\) (\(w_{1}^{12}\)), consider \(\tau_{0,2}\) released at time 6. Retain the stable part \([0, 6]\) from \(w_{1}^{6}\) and convolve the pending part \((6, +\infty)\) with the pmf of \(\mathcal{C}_0\). Combine the results to obtain \(w_{1}^{12}\).

\begin{equation}
    w_1^{12} = f^{[0, 6]}(w_1^6) + \left( f^{(6, +\infty)}(w_1^6) \otimes \mathcal{C}_0 \right)
\end{equation}

When calculating the next busy-window, $\tau_{0,3}$ and $\tau_{1,2}$ are released at 12s. At this point, the sum of probabilities at times greater than 12 (i.e., 13, 14, 15) is 0.000118, which is less than $\epsilon$, satisfying the stop condition. Therefore, the system stops the calculation.

Next, we can use busywindow \( w_i^{r_{i,j}} \) to calculate the backlog \( \mathcal B_{i,j}\). First, sum the stable part of \( w_i^{r_{i,j}} \) to find the probability that the backlog is zero at time \( r_{i,j} \). Then, shift the pending part left by \( r_{i,j} \) units, retain the portion greater than zero, and merge the two parts to obtain \( \mathcal B_{i,j} \).The pmf \( \mathcal B_{i,j} \) is calculated as:

\[
\begin{pmatrix} 0 \\ \sum\limits_{W \leq r_{i,j}} \mathbb{P}(w_i^{r_{i,j}}=W)
\end{pmatrix} \oplus  f^{(0, +\infty)}f_{w_i^{r_{i,j}}}(W_{i,t,k}+r_{i,j})
\]

\begin{figure}[htbp]
    \centering
    \includegraphics[width=0.9\textwidth]{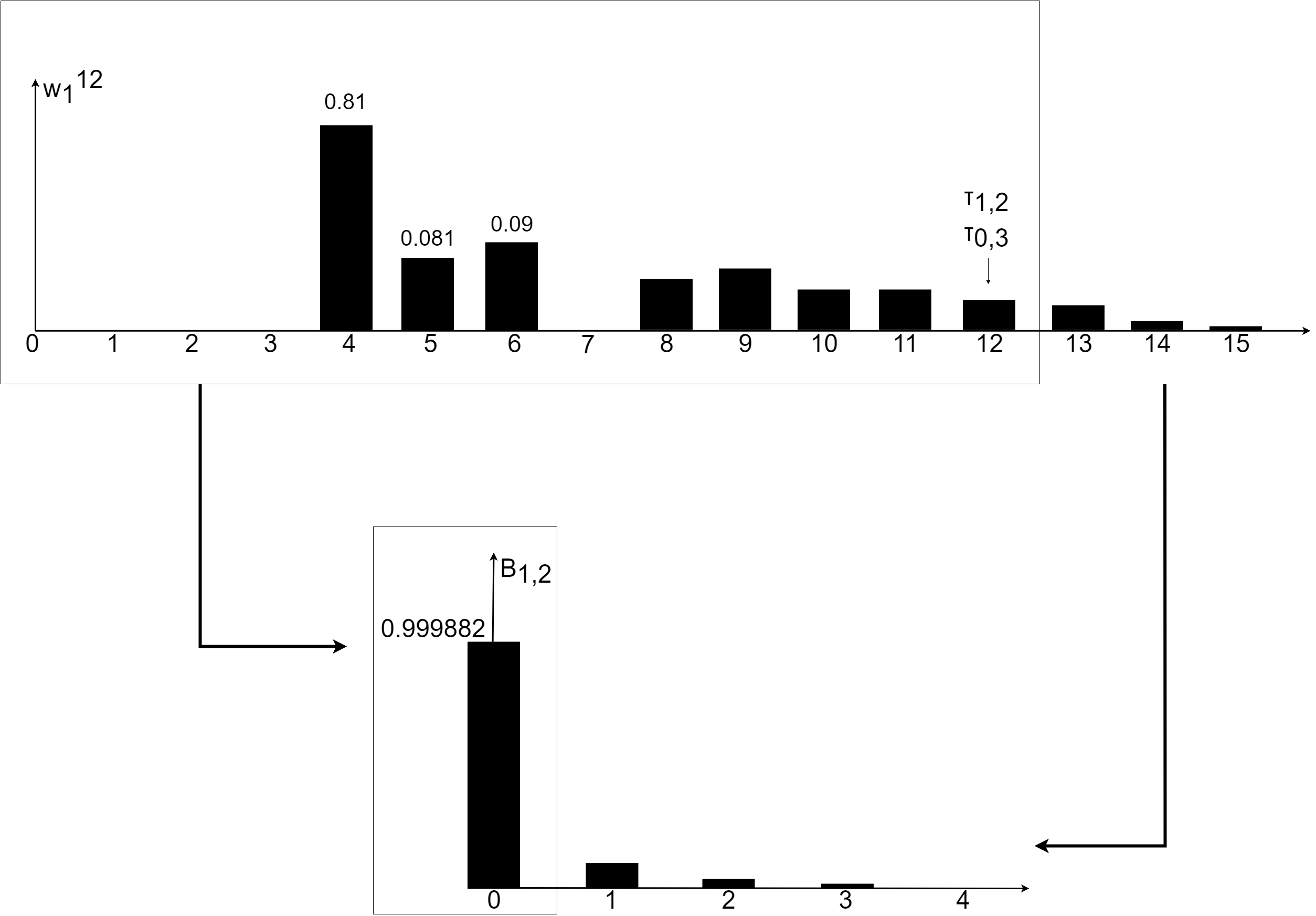}
    \caption{Example for constructing the backlog}
    \label{fig:your_label}
\end{figure}

Fig.4 illustrates how to calculate $\mathcal B_{1,2}$ from $w_{1}^{r_{1,2}}$. Assume we need to compute $B_{1,2}$, which involves finding $w_{1}^{r_{1,2}}$, i.e., $w_{1}^{12}$. First, sum the probabilities in the segment $[0, 12]$ of $w_{1}^{12}$ to obtain the probability $P$ that the backlog is zero. Then, shift the segment $(12, +\infty)$ left by 12 units. Retain the part greater than zero and merge the two parts to obtain $B_{1,2}$.

\begin{equation}
    \mathcal B_{1,2} = \begin{pmatrix} 0 \\ \sum\limits_{W \leq 12} \mathbb{P}(w_{1}^{12} = W) \end{pmatrix} \oplus \left( f^{(0, +\infty)}(w_{1}^{12}) \right)
\end{equation}

$\mathcal{C}_{i,j}^{\text{retrans}}$ represents the cumulative time spent on error retransmissions for frame $\tau_i$. Due to the bus re-arbitrating after each error transmission, it extends the busy window but does not guarantee consecutive successful transmissions thereafter.

\[
\mathcal{C}_{i}^{\text{retrans}} = \mathcal{C}_i - C_i
\]

$\mathcal S_{i,j}^k$ represents the computation of $\mathcal S_{i,j}$ at the time when the $k$-th instance released by $\text{hp}(\tau_i)$ starting from $r_{i,j}$ begins. $\mathcal S_{i,j}^0$ is initialized as:

\[ B_{i,j} \otimes \mathcal{C}_i^{\text{retrans}} \]

Similar to the computation of the busy window, when iterating from \(\mathcal S_{i,j}^k \) to \(\mathcal S_{i,j}^{k+1} \), we retain the segment from \( [0, r_{k-1} - r_{i,j}] \) and proceed with further convolutions using \( (r_{k-1} - r_{i,j}, +\infty) \). If \( k > 0 \), the iterative process can be expressed as:

\[ S_{i,j}^k = f^{[0, r_{k-1} - r_{i,j}]}(S_{i,j}^{k-1}) \oplus \left( f^{(r_{k-1} - r_{i,j}, +\infty)}(S_{i,j}^{k-1}) \otimes \mathcal C_k \right) \]

It is imperative to note that high-priority frames released at time \( r_{i,j} \) will be transmitted prior to \( \tau_{i,j} \), consequently impacting the backlog at \( r_{i,j} \), which corresponds to the zero time instance of the backlog. Thus, during the calculation, the time point \( t=0 \) should be included in the convolution segment.

The stopping condition is when the probability of the next instance falling within \( S_{i,j}^k \) is less than the threshold, which can be expressed as:

\[
\sum_{s_{i,j,n}^k > (r_k-r_{i,j})} \mathbb{P}(S_{i,j}^k=s_{i,j,n}^k) < \epsilon
\]

At this point, once \(\mathcal S_{i,j} = \mathcal S_{i,j}^k \), and \( \tau_{i,j} \) starts transmission successfully, the response time \( \mathcal R_{i,j} \) is computed as:

\[ \mathcal R_{i,j} = \mathcal S_{i,j} \otimes \begin{pmatrix} C_i \\ 1 \end{pmatrix} \]

Therefore, the exceedance function of Probabilistic Worst-Case Response Time (pWCRT) of a frame \(\tau_i\) can be represented as:

\[
F_{R_i}(t) = \max_{j} \left( \sum_{t_0 > t}^{\infty} P(R_{i,j} = t_0) \right)
\]

\section{Experiment} 
\label{sec5}

In this section, we present a comprehensive set of tests designed to evaluate our algorithm from multiple perspectives. We describe the standards or methods used for generating test data, and conduct experiments that measure the accuracy and computational overhead of our algorithm.

\subsection{SAE Benchmark}

\begin{table}[h]
\centering
\caption{SAE CAN Benchmark}
\begin{tabular}{|c|c|c|c|c|c|c|}
\hline
Prio & DLC & C (bits) & E (bits) & T (ms) & D (ms) & R (ms) \\
\hline
1 & 1 & 62 & 13 & 1000 & 5 & 1.416 \\
2 & 2 & 72 & 13 & 5 & 5 & 2.016 \\
3 & 1 & 62 & 13 & 5 & 5 & 2.536 \\
4 & 2 & 72 & 13 & 5 & 5 & 3.136 \\
5 & 1 & 62 & 13 & 5 & 5 & 3.656 \\
6 & 2 & 72 & 13 & 5 & 5 & 4.256 \\
7 & 6 & 112 & 13 & 10 & 10 & 5.016 \\
8 & 1 & 62 & 13 & 10 & 10 & 8.376 \\
9 & 2 & 72 & 13 & 10 & 10 & 8.976 \\
10 & 2 & 72 & 13 & 10 & 10 & 9.576 \\
11 & 1 & 62 & 13 & 100 & 100 & 10.096 \\
12 & 4 & 92 & 13 & 100 & 100 & 19.096 \\
13 & 1 & 62 & 13 & 100 & 100 & 19.616 \\
14 & 1 & 62 & 13 & 100 & 100 & 20.136 \\
15 & 3 & 82 & 13 & 1000 & 1000 & 28.976 \\
16 & 1 & 62 & 13 & 1000 & 1000 & 29.496 \\
17 & 1 & 62 & 13 & 1000 & 1000 & 29.520 \\
\hline
\end{tabular}
\label{table:task_parameters}
\end{table}

\begin{figure}[htbp]
    \centering
    \includegraphics[width=0.8\textwidth]{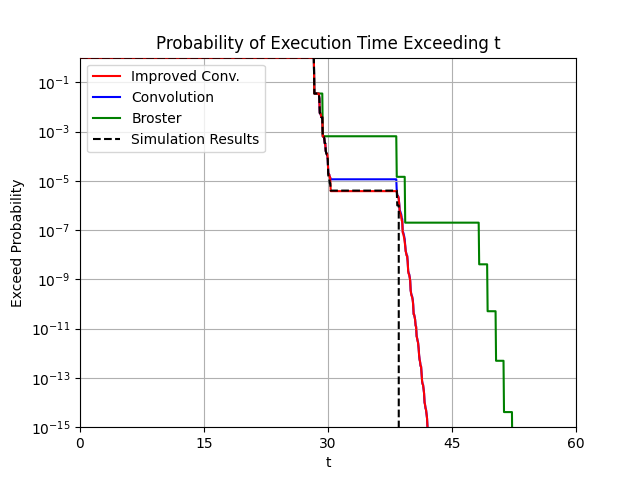}
    \caption{Exceedance functions according to the Improved Convolution Algorithm, Convolution Algorithm, Broster's Approach, and Monte Carlo Simulation with \(10^6\) samples.}
    \label{fig:your_label}
\end{figure}

\begin{table}[h]
    \centering
    \caption{Mean Squared Errors (MSE) compared to Monte Carlo Algorithm}
    \begin{tabular}{|c|c|}
        \hline
        \textbf{Algorithm} & \textbf{Mean Squared Error (MSE)} \\
        \hline
        Improved Convolution Algorithm & 1.4076277135397415e-10 \\
        \hline
        Convolution Algorithm & 1.4818412850130434e-10 \\
        \hline
        Broster's Approach & 4.102067450000266e-06 \\
        \hline
    \end{tabular}
    \label{table:mse_comparison}
\end{table}

To assess the accuracy of the proposed algorithm, we utilize Tindell’s widely published simplification of the Society of Automotive Engineers (SAE) benchmark \citep{tindell1994} \citep{sae1993}. This approach aligns with the data standards used in the Broster's approach and convolution-based algorithm, ensuring transparency and comparability with related work. 

With a bus speed of 125 kbit/s, the bus utilization reaches up to 82.28\%. We employ an error rate of \(\lambda = 10^{-5}\) per bit and set the stopping criterion threshold \(\epsilon\) to \(2.7 \times 10^{-15}\). We focus on calculating the response time probability of the lowest priority frame. This choice is due to the higher system utilization at this frame's priority level, which makes it more susceptible to preemption and generates a longer busy window, thereby better illustrating error injection.

Under the above parameter settings, we calculate the exceedance function for three methods. For each instance of the frame, we take the maximum probability value as the likelihood of the frame's response time exceeding this moment. Because the SAE's lowest priority frame has a period of 1000 ms and the probability \(DFP\) of missing the deadline is 0, selecting the exceedance function better reflects the effectiveness of analyzing the probability distribution.
 To enhance the visibility of the results, we use scientific notation for the vertical axis, choosing the range from \(10^0\) to \(10^{-15}\), as the data in this range are more representative. Research indicates that the worst-case scenario necessarily occurs within the busy window of the critical instant. However, we cannot determine which specific instance will cause the maximum response time \citep{davis2007}. Therefore, we compute the complete busy window and select the maximum value from it to account for the worst-case response time scenario.

To determine the actual error rate, we conduct a Monte Carlo simulation. Specifically, we simulate \(10^6\) critical instant under the given error model and record the probability of the worst-case response time. From the Fig.5, it can be seen that in the [0, 28] range, the exceedance functions obtained by all methods coincide. This range represents the response time obtained in the absence of errors, which is also the minimal necessary response time. In the [30, 40] range, the differences between the three methods become more apparent. Broster's approach introduces significant pessimism, resulting in a greatly increased probability of longer response times. Compared to the convolution-based algorithm, our algorithm accounts for minimal jitter-induced interference, achieving a higher consistency with the results obtained through Monte Carlo simulations. Beyond 40, the overlap in some results occurs because, at this response time, the corresponding frame no longer arrives exactly at the calculated points of the probability distribution function, leading to some convergence in the results.

To evaluate the accuracy of the algorithm’s results compared to the actual results, we use Mean Squared Error (MSE) as the standard. We select 1000 points within the range [0, 60] for evaluation. From the data, it is evident that Broster's approach shows a significant deviation from the Monte Carlo simulation results, whereas the convolution-based algorithm and our method have an MSE difference of approximately \(4 \times 10^{-12}\). This difference is well within our tolerable stopping criterion threshold. Therefore, our method demonstrates minimal error and the highest consistency compared to the Monte Carlo method, showing more optimistic and rigorous results than Broster's approach and the convolution-based algorithm.

\subsection{Randomly Generated Message Sets}

To demonstrate the general accuracy advantage of our algorithm, we randomly generated fifty sets of messages for testing. These task parameters include 50\% system utilization, a fault rate of \(10^{-5}\), and 10 messages per set. The jitter is set to be a random value within the range of \((0, 0.1)\) times the system period. We continue to use a Poisson distribution error retransmission model to calculate the probability distribution of transmission times. The rest of the experimental parameters are consistent with the SAE benchmark, except that we only compute the probability of \(t\) exceeding the deadline in the exceedance function.

We chose the lowest priority frame for calculations to ensure the system utilization is maximized, consistent with the generated settings. Compared to the SAE benchmark, the periods and transmission times of the randomly generated messages exhibit greater variability. The random periods are generally shorter, leading to more frequent arrivals and increased mutual blocking of frames. The larger range in transmission time values results in more frequent \textit{interference} and missed deadlines. To ensure that the experimental results are more pronounced, we plotted each computed instance, thereby avoiding the concentration of deadline miss probabilities around 0\% and 100\%.

As shown in Fig.6, although the three methods are generally consistent in terms of extreme maximum and minimum values, our method yields overall values with an order of magnitude smaller compared to the other two methods. The median is also lower, demonstrating that it has a smaller pessimistic bias and is more accurate.

\begin{figure}[htbp]
    \centering
    \includegraphics[width=0.8\textwidth]{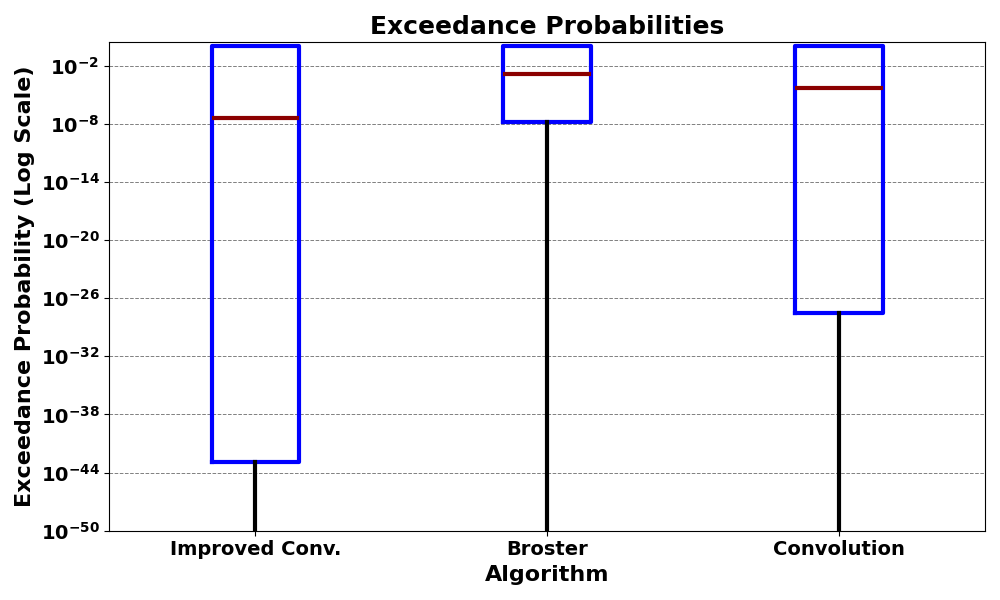}
    \caption{Boxplot of deadline miss probabilities at 50\% utilization for the Improved Convolution Algorithm, Convolution Algorithm, and Broster's Approach.}
    \label{fig:your_label}
\end{figure}

At higher utilization levels, the probability of the worst-case response time exceeding the deadline becomes more concentrated around 1. Conversely, at lower utilization levels, this probability becomes more concentrated around 0. In both cases, the experimental effects are less pronounced. Therefore, we chose to generate random tasks with 50\% utilization to achieve more noticeable results.

In a similar manner to the 50\% utilization setup, we computed random message sets for different utilization levels. The experimental results shown in Fig.7 were broadly consistent with previous findings, causing the probability of missing deadlines to approach the two extremes. Consequently, the differences in results were no longer sufficiently significant.

\begin{figure}[htbp]
    \centering
    \includegraphics[width=0.8\textwidth]{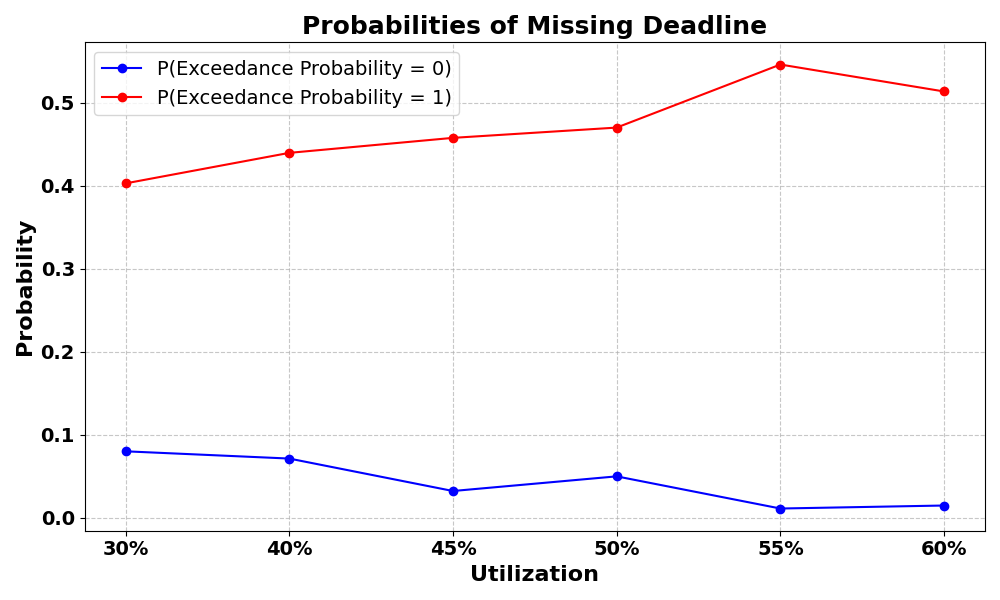}
    \caption{Change in the proportion of pWCRT exceedance probabilities being 1 and 0 with varying utilization levels.}
    \label{fig:utilization}
\end{figure}

\subsection{Computational Overhead of Convolution Operations}

Compared to convolution methods that require starting from time zero for each iteration, our approach only needs to calculate the busy window once and then compute the backlog, significantly reducing computational overhead. Using the random message set with 50\% utilization from the previous experiment, we measured the time required for iterating through the lowest priority task using both convolution algorithms. As shown in the Fig.8, our method consistently incurs lower time overhead than the original convolution method.

When the number of instances for calculating the busy window increases, the original convolution algorithms must start iteration from scratch each time, resulting in more frequent recalculations of the same window. The overall number of instances tends to increase with higher system utilization, and message sets with high utilization naturally lead to longer windows. Therefore, the advantage of our improved method in terms of computational overhead becomes more pronounced under high utilization conditions. We calculated the average and maximum time overheads of the two algorithms across 50 sets with 10 messages each, within the 30\% to 60\% utilization range. As shown in Figs.9 and 10, the time overhead increases significantly at higher utilization, especially in extreme cases that produce the maximum overhead.

Additionally, under the same utilization conditions, an increase in the number of messages further amplifies the computational overhead of convolution, making the time difference between the original and improved convolution algorithms increasingly apparent. In practical CAN bus communication environments, the number of ECUs can reach dozens to hundreds, with corresponding messages numbering in the hundreds or even thousands. In such scenarios, the advantage of our algorithm becomes significant, making it more suitable for industrial environments. Similarly, at 60\% utilization, we conducted 50 experiments on 5 to 30 messages. As shown in Figs.11 and 12, the increase in computational overhead generally follows a consistent trend with the increase in the number of messages, although some extreme cases are influenced by specific message sets.

\begin{figure}[htbp]
    \centering
    \includegraphics[width=0.8\textwidth]{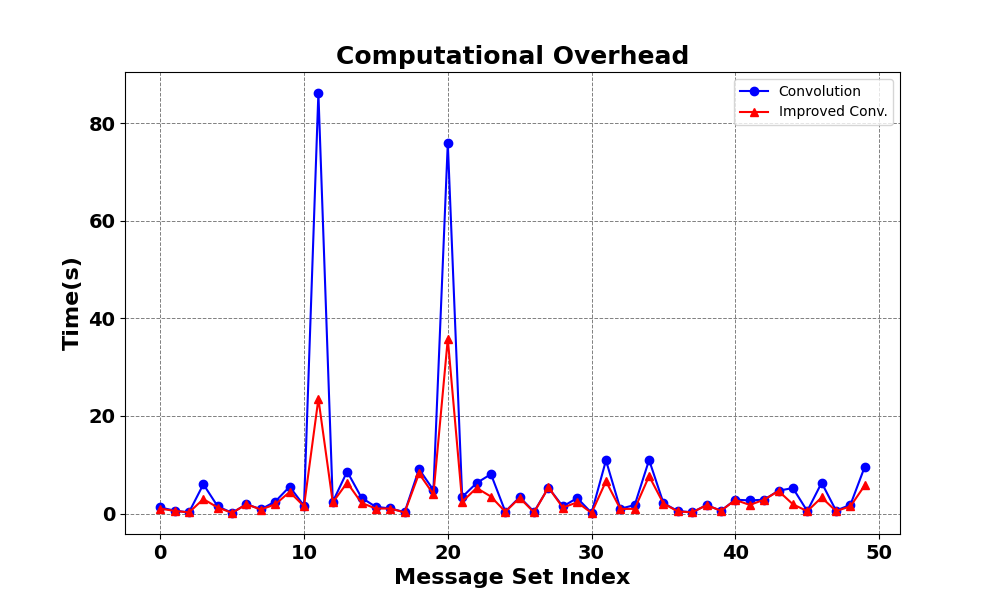}
    \caption{Computational time overhead of the Improved Convolution Algorithm and Convolution Algorithm at 50\% system utilization.}
    \label{fig:utilization}
\end{figure}

\begin{figure}[h!]
    \centering
    \begin{minipage}[b]{0.45\textwidth} 
        \centering
        \includegraphics[width=\textwidth]{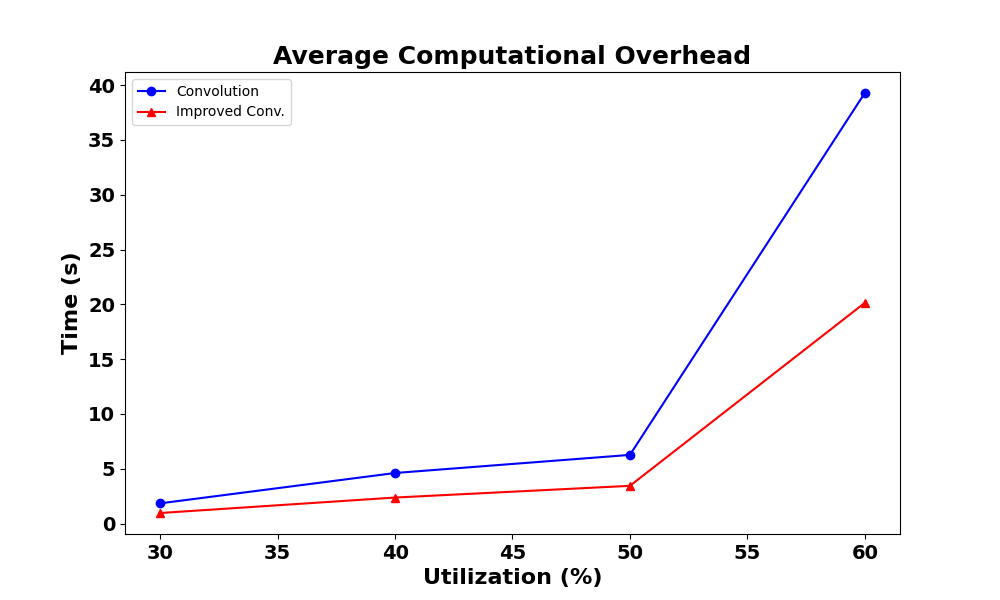}
        \caption{Average computational overhead of 50 sets with 10 messages under different utilizations.}
        \label{fig:speed}
    \end{minipage}
    \hfill
    \begin{minipage}[b]{0.45\textwidth} 
        \centering
        \includegraphics[width=\textwidth]{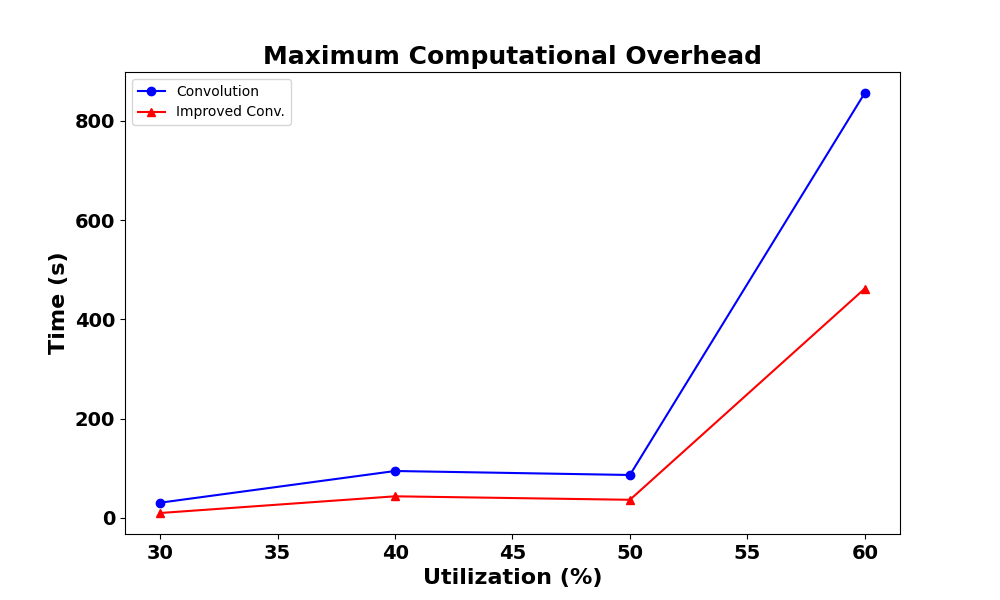}
        \caption{Maximum computational overhead of 50 sets with 10 messages under different utilizations.}
        \label{fig:speed_max_1}
    \end{minipage}    
    
    \vspace{0.5cm} 
    
    \begin{minipage}[b]{0.45\textwidth} 
        \centering
        \includegraphics[width=\textwidth]{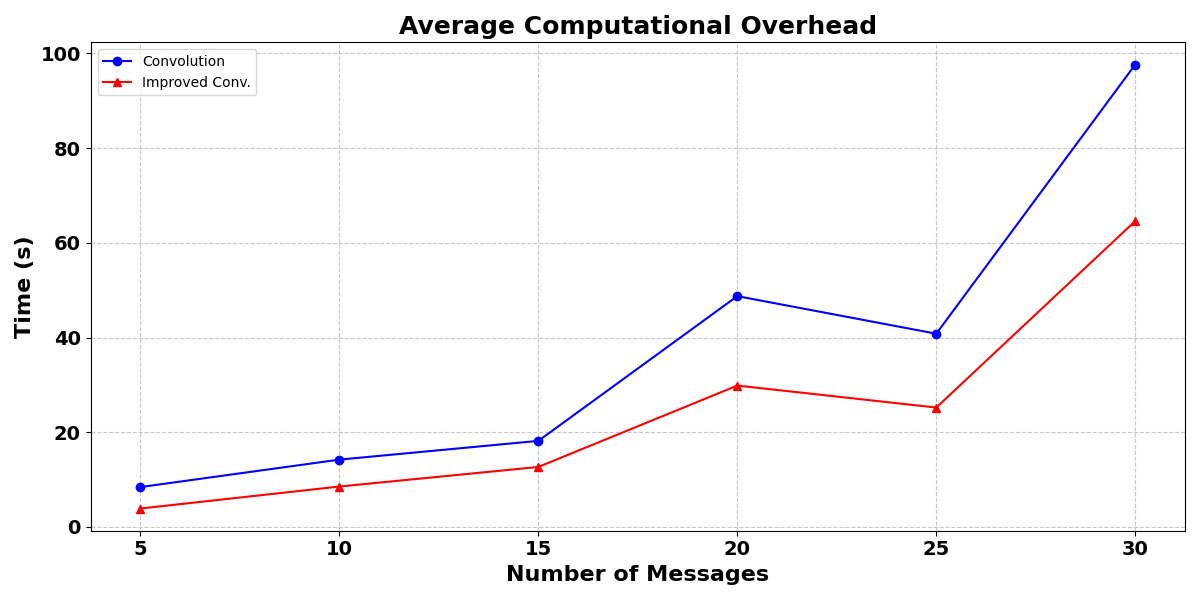}
        \caption{Average computational overhead of 50 sets at 60\% utilization with different message counts.}
        \label{fig:speed_max_2}
    \end{minipage}
    \hfill
    \begin{minipage}[b]{0.45\textwidth} 
        \centering
        \includegraphics[width=\textwidth]{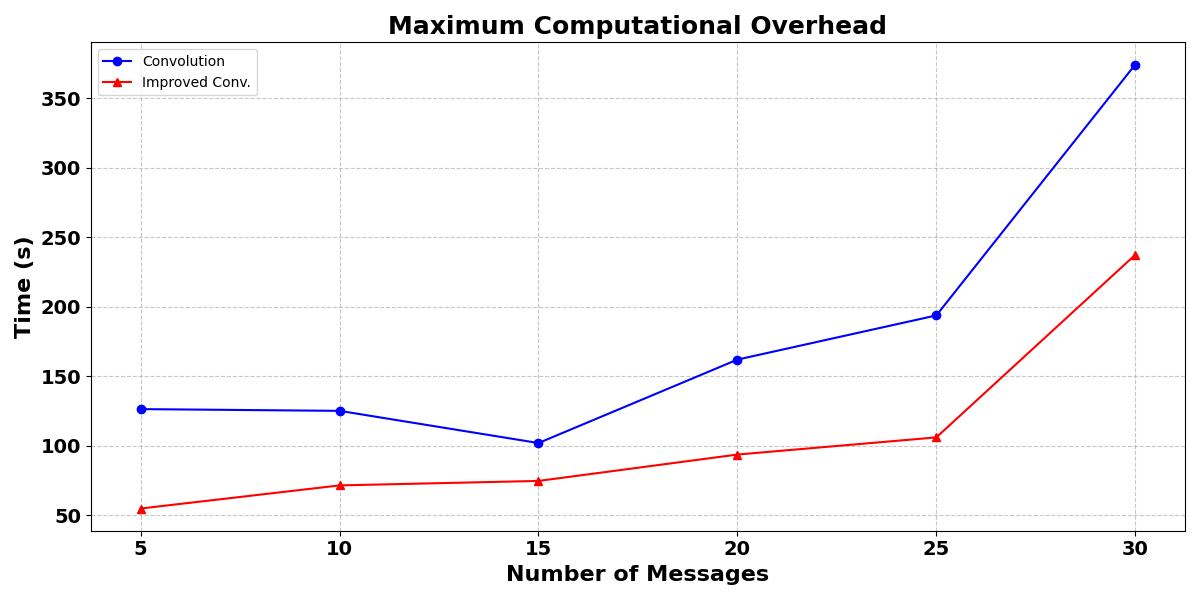}
        \caption{Maximum computational overhead of 50 data sets at 60\% utilization with different message counts.}
        \label{fig:speed_max_3}
    \end{minipage}
\end{figure}

\subsection{Discussion}

In the SAE Benchmark experiments, we used Mean Squared Error (MSE) to assess the discrepancy between algorithmic results and simulation outcomes. However, this approach tends to compress the scale of errors and is less sensitive to low-probability events, making it an imperfect evaluation metric. Therefore, identifying a more effective metric to assess algorithm accuracy remains a valuable area for discussion. Additionally, although we have made efforts to reduce computational overhead, the convolution method still incurs significant time costs under high utilization and with a large number of messages. Finding more efficient ways to reduce the time overhead of analysis iterations remains a future challenge \citep{markovic2021} \citep{bozhko2021} \citep{vonderbruggen2018}.
.

\section{Conclusion and Future Work} 
\label{sec6}

Formal verification is essential for ensuring the reliability and safety of real-time systems. In this study, we enhance a convolution-based probabilistic response time analysis method by incorporating busy window and backlog techniques. Our approach delivers a probabilistic distribution of system response times, integrating an error retransmission mechanism. By comparing these probability distributions with relative deadlines, we can derive the probability of deadline misses, facilitating the prediction and assurance of system safety-critical aspects. Our algorithm reduces pessimism without introducing optimism, aligning results more closely with simulation outcomes. It also minimizes the impact of new tasks arriving precisely at the end of the busy window, thereby shortening the computational window. Furthermore, by reusing busy window calculations to determine the backlog, we significantly reduce redundant convolution operations, with this reduction being particularly pronounced at higher system utilization levels and with an increasing number of messages. However, our algorithm still faces challenges such as time overhead, and further improvements and research on analysis methods remain areas for future work.







\begin{thebibliography}{00}

\bibitem[Axer \& Ernst(2013)]{axer2013}
  Philip Axer and Rolf Ernst,
  \textit{Stochastic response-time guarantee for non-preemptive, fixed-priority scheduling under errors},
  Proceedings of the 50th Annual Design Automation Conference,
  2013.

\bibitem[Broster \textit{et al.}(2002)]{broster2002}
  Ian Broster, Alan Burns, and Guillermo Rodriguez-Navas,
  \textit{Probabilistic analysis of CAN with faults},
  Proceedings of the 23rd IEEE Real-Time Systems Symposium (RTSS 2002),
  IEEE,
  2002.

\bibitem[Tindell \textit{et al.}(1994)]{tindell1994}
  Ken Tindell and Alan Burns,
  \textit{Guaranteeing message latencies on control area network (CAN)},
  Proceedings of the 1st International CAN Conference,
  Citeseer,
  1994.

\bibitem[SAE(1993)]{sae1993}
  Society of Automotive Engineers,
  \textit{Class C Application Requirement Considerations},
  SAE J2056,
  Society of Automotive Engineers,
  1993.


\bibitem[Díaz \textit{et al.}(2002)]{diaz2002}
  José Luis Díaz, et al.,
  \textit{Stochastic analysis of periodic real-time systems},
  23rd IEEE Real-Time Systems Symposium, RTSS 2002,
  IEEE,
  2002.

\bibitem[Broster et al.(2004)]{broster2004}
  Ian Broster, Alan Burns, and Guillermo Rodríguez-Navas,
  \textit{Comparing real-time communication under electromagnetic interference},
  Proceedings of the 16th Euromicro Conference on Real-Time Systems (ECRTS 2004),
  IEEE,
  2004.

\bibitem[Davis \textit{et al.}(2007)]{davis2007}
  Robert I. Davis, et al.,
  \textit{Controller Area Network (CAN) schedulability analysis: Refuted, revisited and revised},
  Real-Time Systems,
  vol. 35,
  pp. 239--272,
  2007.

\bibitem[Liu and Layland(1973)]{liu1973}
  Chung Laung Liu and James W. Layland,
  \textit{Scheduling algorithms for multiprogramming in a hard-real-time environment},
  Journal of the ACM (JACM),
  vol. 20,
  no. 1,
  pp. 46--61,
  1973.

\bibitem[IEC(2010)]{iec2010}
  International Electrotechnical Commission (IEC),
  \textit{Functional safety of electrical / electronic / programmable electronic safety-related systems (ed2.0)},
  IEC,
  2010.

\bibitem[ISO(2000)]{iso2000}
  International Organization for Standardization (ISO),
  \textit{ISO/FDIS 26262: Road vehicles – Functional safety},
  ISO,
  2000.

\bibitem[Bozhko \textit{et al.}(2023)]{bozhko2023}
  Sergey Bozhko, et al.,
  \textit{What Really is pWCET? A Rigorous Axiomatic Proposal},
  2023 IEEE Real-Time Systems Symposium (RTSS),
  IEEE,
  2023.

\bibitem[Chow \& Teicher(2012)]{chow2012}
  Yuan Shih Chow and Henry Teicher,
  \textit{Probability theory: independence, interchangeability, martingales},
  Springer Science \& Business Media,
  2012.

\bibitem[Bini \& Buttazzo(2005)]{bini2005}
  Enrico Bini and Giorgio C. Buttazzo,
  \textit{Measuring the performance of schedulability tests},
  Real-Time Systems,
  vol. 30,
  no. 1,
  pp. 129--154,
  2005.

\bibitem[Marković \textit{et al.}(2021)]{markovic2021}
  Filip Marković, Alessandro Vittorio Papadopoulos, and Thomas Nolte,
  \textit{On the convolution efficiency for probabilistic analysis of real-time systems},
  33rd Euromicro Conference on Real-Time Systems (ECRTS 2021),
  Schloss Dagstuhl-Leibniz-Zentrum für Informatik,
  2021.

\bibitem[Bosch(1991)]{bosch1991}
  Specification, C. A. N.,
  \textit{Bosch},
  Robert Bosch GmbH, Postfach 50,
  1991.

\bibitem[Sebastian \& Ernst(2009)]{sebastian2009}
  Maurice Sebastian and Rolf Ernst,
  \textit{Reliability analysis of single bus communication with real-time requirements},
  2009 15th IEEE Pacific Rim International Symposium on Dependable Computing,
  IEEE,
  2009.

\bibitem[Bozhko \textit{et al.}(2021)]{bozhko2021}
  Sergey Bozhko, Georg von der Brüggen, and Björn Brandenburg,
  \textit{Monte Carlo response-time analysis},
  IEEE 42nd Real-Time Systems Symposium,
  IEEE,
  2021.

\bibitem[Leen \& Heffernan(2002)]{leen2002}
  Gabriel Leen and Donal Heffernan,
  \textit{TTCAN: a new time-triggered controller area network},
  \textit{Microprocessors and Microsystems},
  vol. 26, no. 2,
  pp. 77--94,
  2002.

\bibitem[Maxim \& Cucu-Grosjean(2013)]{maxim2013}
  Dorin Maxim and Liliana Cucu-Grosjean,
  \textit{Response time analysis for fixed-priority tasks with multiple probabilistic parameters},
  Proceedings of the 2013 IEEE 34th Real-Time Systems Symposium (RTSS),
  IEEE,
  2013.

\bibitem[Woodbury \& Shin(1988)]{Woodbury1988}
  Michael H. Woodbury and Kang G. Shin,
  \textit{Evaluation of the Probability of Dynamic Failure and Processor Utilization for Real-Time Systems},
  Proceedings of the Real-Time Systems Symposium,
  IEEE Computer Society,
  1988.

\bibitem[Tia \textit{et al.}(1995)]{Tia1995}
  Tia, T-S., G. C. Buttazzo, and H. A. B. S. M. Ho,
  \textit{Probabilistic Performance Guarantee for Real-Time Tasks with Varying Computation Times},
  Proceedings of the Real-Time Technology and Applications Symposium,
  IEEE,
  1995.

\bibitem[Gardner \& Liu(1999)]{Gardner1999}
  Gardner, Mark K., and Jane W. S. Liu,
  \textit{Analyzing Stochastic Fixed-Priority Real-Time Systems},
  International Conference on Tools and Algorithms for the Construction and Analysis of Systems,
  Springer Berlin Heidelberg,
  1999.

\bibitem[Diaz \textit{et al.}(2004)]{Diaz2004}
  Diaz, Jose Luis, et al.,
  \textit{Pessimism in the Stochastic Analysis of Real-Time Systems: Concept and Applications},
  25th IEEE International Real-Time Systems Symposium,
  IEEE,
  2004.

\bibitem[Tanasa \textit{et al.}(2015)]{Tanasa2015}
  Tanasa, Bogdan, et al.,
  \textit{Probabilistic Response Time and Joint Analysis of Periodic Tasks},
  2015 27th Euromicro Conference on Real-Time Systems,
  IEEE,
  2015.

\bibitem[Ivers \& Ernst(2009)]{Ivers2009}
  Ivers, Matthias, and Rolf Ernst,
  \textit{Probabilistic Network Loads with Dependencies and the Effect on Queue Sojourn Times},
  Quality of Service in Heterogeneous Networks: 6th International ICST Conference on Heterogeneous Networking for Quality, Reliability, Security and Robustness, QShine 2009 and 3rd International Workshop on Advanced Architectures and Algorithms for Internet Delivery and Applications, AAA-IDEA 2009, Las Palmas, Gran Canaria, November 23-25, 2009 Proceedings 6,
  Springer Berlin Heidelberg,
  2009.

\bibitem[Tanasa et al.(2013)]{Tanasa2013}
  Tanasa, Bogdan, et al.,
  \textit{Probabilistic Timing Analysis for the Dynamic Segment of FlexRay},
  25th Euromicro Conference on Real-Time Systems,
  IEEE,
  2013.

\bibitem[von der Brüggen \textit{et al.}(2018)]{vonderbruggen2018}
  G. von der Brüggen, N. Piatkowski, K. H. Chen, \textit{et al.},
  \textit{Efficiently Approximating the Probability of Deadline Misses in Real-Time Systems},
  30th Euromicro Conference on Real-Time Systems (ECRTS 2018),
  Schloss-Dagstuhl-Leibniz Zentrum für Informatik,
  2018.

\bibitem[Bernat \textit{et al.}(2003)]{bernat2003}  
  Guillem Bernat, Antoine Colin, and Stefan Petters,  
  \textit{pwcet: A tool for probabilistic worst-case execution time analysis of real-time systems},  
  Proceedings of the 3rd International Workshop on Worst-Case Execution Time Analysis (WCET 2003),  
  2003.

\bibitem[Davis \& Cucu-Grosjean(2019)]{davis2019}  
  Robert Ian Davis and Liliana Cucu-Grosjean,  
  \textit{A survey of probabilistic timing analysis techniques for real-time systems},  
  LITES: Leibniz Transactions on Embedded Systems,   
  2019, pp. 1-60.

\bibitem[Diaz \textit{et al.}(2004)]{diaz2004}  
  Jose Luis Diaz, Diego F. Garcia, Mario Garcia, Angel Gonzalez, Ramon Orti, and Javier Zamorano,  
  \textit{Pessimism in the stochastic analysis of real-time systems: Concept and applications},  
  25th IEEE International Real-Time Systems Symposium (RTSS),  IEEE, 
  2004.

\end{thebibliography}



\appendix
\section{Pseudo Code}
Algorithm~\ref{alg:random_busy_window} presents the pseudocode implementation for the random busy window calculation. As previously described, the algorithm starts by iteratively computing the maximum blocking time each frame experiences due to lower-priority frames. Then, it initializes the state of each busy window by adding itself and higher-priority frames to the waiting queue, with all arriving at time 0. 

The iteration process begins by sorting frames based on their arrival times and priorities to determine the next convolution frame $\tau$ and its arrival time $t$. The interval \([0, t]\) is designated as the stable part and saved, while the interval \((t, +\infty)\) is considered as the pending part, which is convolved with $\tau$. The two parts are then merged to form a new busy window. If the probability of the pending part is below a threshold, the loop terminates and the iteration stops.

Finally, the arrival times of the frames involved in the convolution are updated by period and added back to the waiting queue, and the process continues.

\begin{algorithm}
\label{alg:random_busy_window}
\caption{Computing Busy Window for Instances of Frames $F$}
\SetAlgoLined
\KwIn{Frames $F = (\tau_1, \tau_2, \ldots, \tau_n)$, where each $\tau_i = (C_i, D_i, T_i, p_i)$, pmf of $\mathcal{C}_i$, Threshold $\epsilon$}
\KwOut{Longest Busy Window $w_{i,j}$}

\textbf{Initialization:}\\
$B_n = 0$\;
\For{each frame $\tau_i \in F$ in reverse order}{
    $B_i = \max(B_{i+1}, C_i + E)$\;
    Initialize busy window $w_i^0$ as $(B_i, 1)$\;
}
$Queue \gets []$\;
\For{each frame $\tau_i \in F$}{
    Generate an instance of $\tau_i$ at $t=0$\;
    $Queue.\text{append}((\tau_i, 0))$\;
}

\textbf{Iterate Busy-Window:}\\
\While{Queue is not empty}{
    Sort $Queue$ by release time and priority\;
    $(\tau, t) \gets Queue.pop(0)$\;
    $w_{stable} \gets f^{[0, t]}(w_i^t)$\;
    $w_{pending} \gets f^{(t, +\infty)}(w_i^t)$\;
    \If{$\sum P(w_{pending}) < \epsilon$}{
        \textbf{break}\;
    }
    $w_{pending} \gets w_{pending} \otimes \mathcal{C}_\tau$\;
    $w_i^t \gets $ Merge($w_{stable}$, $w_{pending}$)\;
    
    Generate the next instance of $\tau$ at $t + T_\tau$\;
    $Queue.\text{append}((\tau, t + T_\tau))$\;
}

\end{algorithm}
\end{document}